\documentclass[aps,prc,twocolumn,floatfix,superscriptaddress,nofootinbib]{revtex4-1}
\usepackage{xcolor}
\usepackage{graphicx}
\usepackage{dcolumn}

\begin{document}

\title{Origin of fine structure of the giant dipole resonance in $sd$-shell nuclei}

\author{R. W. Fearick}
\affiliation{Department of Physics, University of Cape Town, Rondebosch 7700, South Africa}
\author{B. Erler}
\affiliation{Institut f\"ur Kernphysik, Technische Universit\"at Darmstadt,
D-64289 Darmstadt, Germany}
\author{H. Matsubara}
\affiliation{Research Center for Nuclear Physics, Osaka University, Ibaraki, 
Osaka 567-0047, Japan}
\affiliation{Tokyo Women's Medical University, 8-1 Kawada-cho, Shinjuku-ku, Tokyo 162-8666, Japan}
\author{\mbox{P. von Neumann-Cosel}}\email{vnc@ikp.tu-darmstadt.de}
\affiliation{Institut f\"ur Kernphysik, Technische Universit\"at Darmstadt,
D-64289 Darmstadt, Germany}
\author{A. Richter}
\affiliation{Institut f\"ur Kernphysik, Technische Universit\"at Darmstadt,
D-64289 Darmstadt, Germany}
\author{R. Roth}
\affiliation{Institut f\"ur Kernphysik, Technische Universit\"at Darmstadt,
D-64289 Darmstadt, Germany}
\author{A. Tamii}
\affiliation{Research Center for Nuclear Physics, Osaka University, Ibaraki, 
Osaka 567-0047, Japan} 

\date{\today}

\begin{abstract}
A set of high resolution zero-degree inelastic proton scattering data on $^{24}$Mg, $^{28}$Si, $^{32}$S, and $^{40}$Ca 
provides new insight into  the long-standing puzzle of the origin of fragmentation of the Giant Dipole Resonance (GDR) in $sd$-shell nuclei. 
Understanding is achieved by comparison with Random Phase Approximation (RPA) calculations for deformed nuclei using for the first time a realistic nucleon-nucleon interaction derived from the Argonne V18 potential with the unitary correlation operator method and supplemented by a phenomenological three-nucleon contact interaction.
A wavelet analysis allows to extract significant scales both in the data and calculations characterizing the fine structure of the GDR. 
The fair agreement for scales in the range of a few hundred keV supports the surmise that the fine structure arises from ground-state deformation driven by $\alpha$ clustering. 
\end{abstract}

\maketitle

\section{Introduction}

%
\begin{figure*}
\includegraphics[width=15cm]{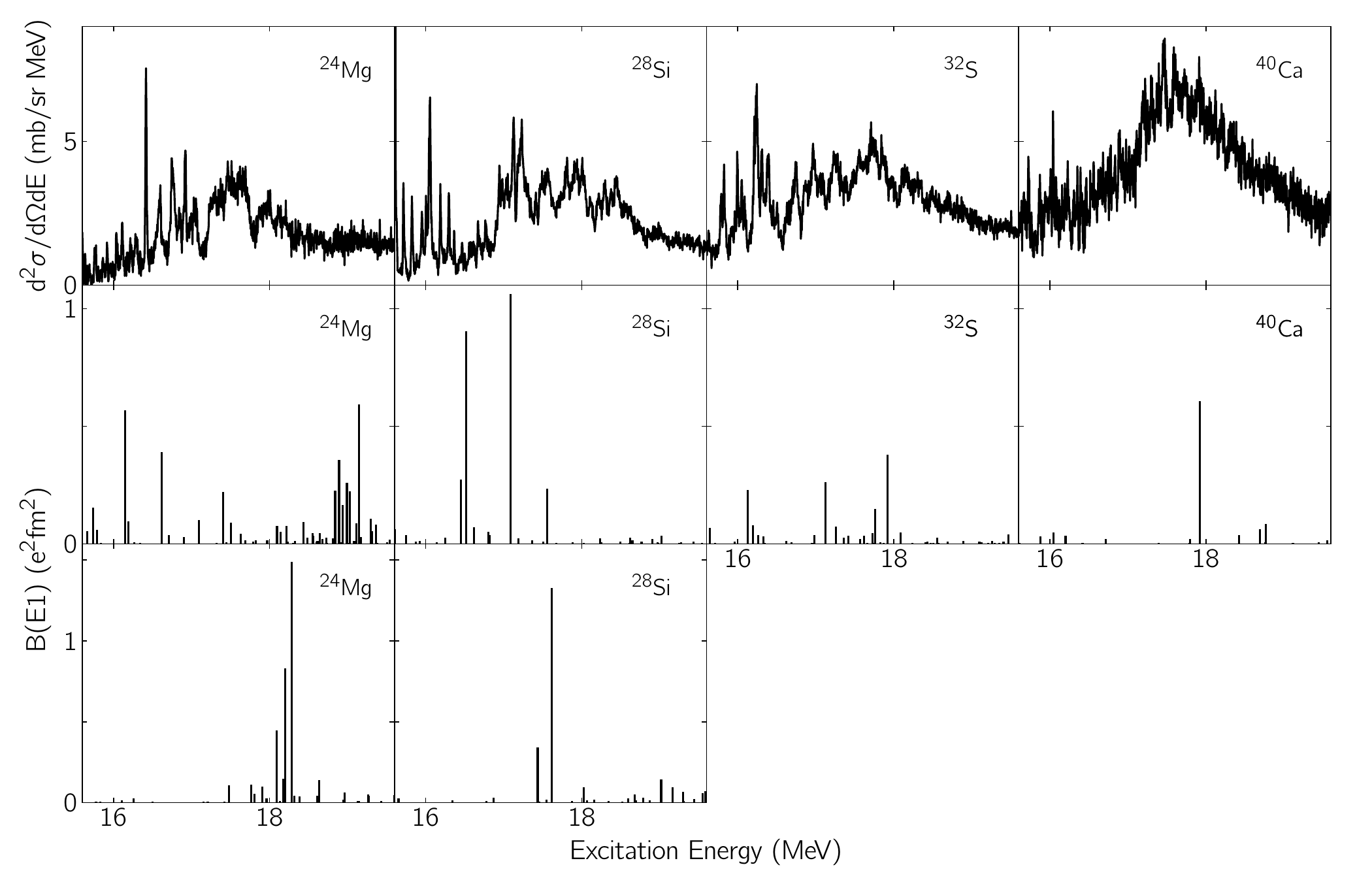}
\caption{Top row: High-resolution ($20-30$ keV FWHM) spectra of the $(p,p^\prime)$ reaction at $E_0=295$ MeV and $\theta_{lab}=0.4^\circ$ for $^{24}$Mg, $^{28}$Si, $^{32}$S, and $^{40}$Ca.  
The excitation energy range shown encompasses the GDR.
Middle row: Theoretical $B(E1)$ strength distributions calculated in a deformed-basis RPA with the UCOM interaction \cite{erl14}.
Bottom row: Theoretical $B(E1)$ strength distributions  in $^{24}$Mg and $^{28}$Si computed in a HFB+QRPA approach with the D1S Gogny force \cite{per08}.}
\label{fig:data}
\end{figure*}

The isovector giant dipole resonance is the best known of the fundamental collective excitations of the nucleus \cite{har01}. 
A giant resonance can be understood macroscopically as a bulk nuclear vibration, and microscopically in terms of coherent particle-hole excitations. 
The gross properties  of the GDR such as the centroid in energy of the excitations and the strength in terms of sum rules are well understood. 
Less well understood are, however, the details of decay processes of the resonance. 
Various contributions to the width of the giant resonances have been identified \cite{har01,bbb98}: direct decay out of the continuum leading to an escape width, coupling to two-particle two-hole states ($2p2h$) and then to many-particle many-hole ($npnh$) states giving rise to a spreading width, and fragmentation of the elementary $1p1h$ states that form the resonance called Landau damping.
These processes contribute to the total width of the resonance and manifest themselves experimentally by different structures in the excitation spectra. 

A fragmentation of the GDR in $p$- and $sd$-shell nuclei on the scale of several MeV is long-established and has already early been interpreted as configurational splitting \cite{era86}.
Recently, it has been argued that the strength distribution of the GDR in $^{12}$C and $^{16}$O reveals information on the role of different $\alpha$-cluster configurations \cite{he14}.
However, the observation of finer structures of the GDR in light nuclei on the scale of several hundred keV remains a puzzle.
In general, the physical origin of this kind of structure must be related to the existence of complex configurations, different time scales in decay processes, or the removal of the angular-momentum substate degeneracy due to deformation.
Taking $^{28}$Si as an example of an $sd$-shell nucleus, structure on the finest scales was observed in reaction cross sections \cite{sin65} and identified with Ericson fluctuations \cite{eri60}.
These are essentially a manifestation of the spreading width.

Fine structure of the GDR has also been observed in heavy nuclei \cite{iwa12,pol14} and in other giant resonances such as the isoscalar giant quadrupole resonance (GQR)  \cite{she04,she09}, the Gamow-Teller resonance \cite{kal06}, or the magnetic quadrupole resonance \cite{vnc99,kal07}. 
Some progress has been made in the understanding of the fine structure by comparison between experiments and theoretical calculations of the distribution of resonance strength.
In the case of the GQR it has been demonstrated that the fine structure has its origin in the coupling of the $1p1h$ excitations that constitute the resonance to low-lying surface vibrations \cite{she04,she09}, a mechanism discussed in Ref.~\cite{ber83}.
However, in recent studies of lighter nuclei ($^{28}$Si, $^{40}$Ca) it was shown that Landau damping plays a role in the formation of fine structure \cite{usm11,usm16}. 
The importance of Landau damping was also demonstrated for the GDR in $^{208}$Pb \cite{pol14}. 

Here we turn attention to the GDR in lighter nuclei with equal proton and neutron numbers $Z$ and $N$, respectively. 
In the $sd$ shell these nuclei are deformed and according to a recent theoretical study \cite{erl14} a key driver of deformation is the underlying $\alpha$-cluster structure. 
Here we focus on the nuclei $^{24}$Mg, $^{28}$Si, $^{32}$S and $^{40}$Ca, which allows to compare nuclei with prolate deformed ($^{24}$Mg,$^{32}$S), oblate deformed ($^{28}$Si), and spherical  ground states ($^{40}$Ca).
Does this structural feature manifest itself in the fragmentation of the GDR?
An answer to this question has become possible by the confluence of two advances: 
(i) The recent availability of high-resolution zero degree inelastic proton scattering data from a series of light nuclei \cite{tam09,mat15} permits fine structure in the spectra to be resolved while providing a high degree of selectivity towards $1^-$ states,  
and (ii) the availability of microscopic calculations of the GDR strength using the random phase approximation (RPA) on top of a deformed ground state with modern nucleon-nucleon interactions. 
These calculations do not yet include coupling to the continuum or to more complex configurations but probe the effect of deformation on the fine structure of the GDR in $sd$-shell nuclei.
As shown below, a detailed comparison yields good agreement with experiment, which leads us to the conclusion that deformation plays a key role in the formation of fine structure in these $sd$-shell nuclei.  

A central part of this work is a quantitative characterization of scales of fragmentation in the GDR region. 
Various measures have been proposed in the past, viz.\ averaging of spectra at various scales \cite{sin65}, Fourier analysis \cite{ric74}, correlation analysis \cite{kil87}, the entropy index method \cite{lac99,lac00}, local scaling dimension \cite{aib99,aib11}, and wavelet analysis \cite{she08}. 
We have chosen wavelet analysis as it offers a quantification of the energy scales of fine structures while resolving the strength of fine structures in both excitation energy and energy scale.
Thus structures can be localized within the excitation energy region of the GDR.
The wavelet analysis of the experimental spectra and corresponding theoretical strength distributions then permits us to make comparisons of the derived energy scales in different nuclei. 
Given the complexity of nuclear behavior, such comparisons are necessarily of semi-quantitative nature. 
We do, however, expect some insight into the physical origin of the scales of structures.

\section{Experiment}

Measurements of inelastic proton scattering at high resolution and at forward angles including $0^\circ$ have only recently become feasible \cite{tam09,nev11}. 
The present data were taken at RCNP, Osaka, Japan  with the Grand Raiden magnetic spectrometer \cite{fuj99} with 295 MeV proton beams.
Dispersion-matching techniques were applied to achieve high energy-resolution of the order 20 to 30 keV full width at half maximum (FWHM) at angles near zero degree \cite{tam09}.
Targets consisted of isotopically enriched thin foils with areal densities of a few mg/cm$^2$. 
The spectrometer placed at $0^\circ$ covered an angular accpetance of $\pm 2.5^\circ$.
Additional data were taken with the spectrometer placed at larger angles covering an angular range up to $15^\circ$. 
The spectra analyzed in the present work correspond to a mean scattering angle of $0.4^\circ$, where the cross sections for excitation of $1^-$ states by relativistic Coulomb excitation dominate \cite{ber88}.
The momentum acceptance of the spectrometer permitted to cover a range of roughly $5-25$ MeV in excitation energy. 
Spectra of the GDR between 14 and 24 MeV after subtraction of instrumental background are shown in the top row of  Fig.~\ref{fig:data}. 
A considerable amount of fine structure is observed.

\begin{figure}[t]
\includegraphics[width=8.5cm]{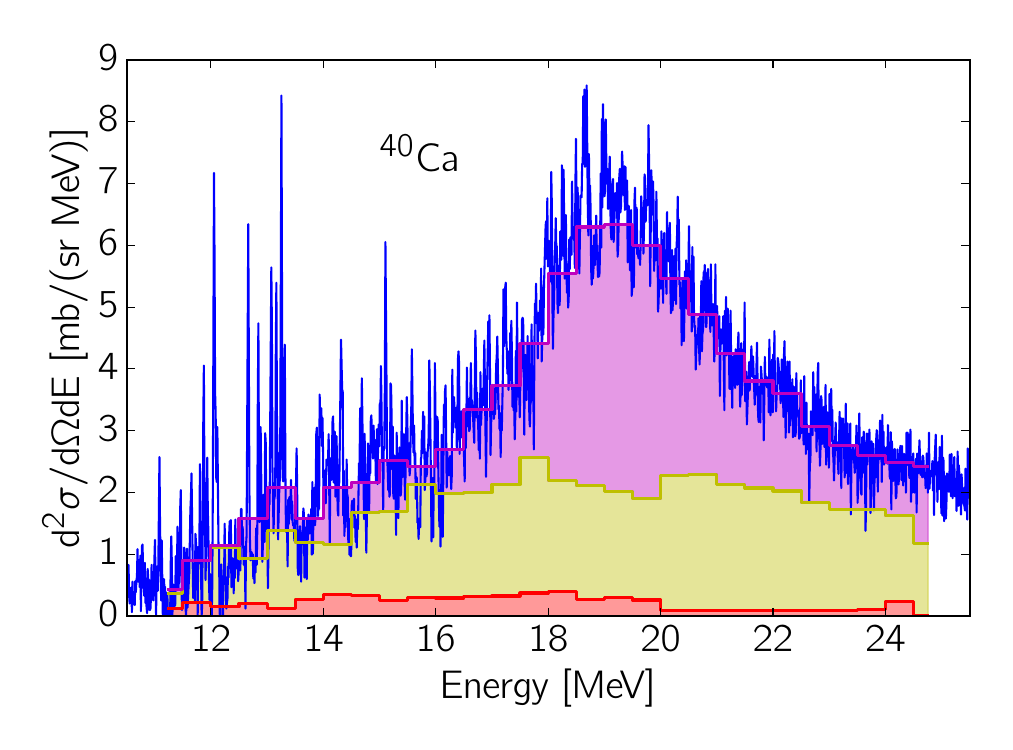}
\caption{Multipole decomposition of the $^{40}$Ca$(p,p^\prime)$ cross sections at $\Theta_{\rm lab} = 0^\circ -2.5^\circ$  (blue histogram) for 200 keV bins.
Purple: E1.
Green: Quasifree scattering.
Red: Isoscalar giant resonances (E0+E2+E3). }
\label{fig:mda}
\end{figure}

Before starting the analysis we demonstrate that the spectra in Fig.~\ref{fig:data} are indeed dominated by relativistic Coulomb excitation of E1 transitions and thus the fine structure visible is related to the GDR.
Figure \ref{fig:mda} displays by way of example a multipole decomposition analysis (MDA) of the angular distribution for the case of $^{40}$Ca.
Details of the MDA approach are described in the analysis of comparable data on heavier nuclei \cite{tam11,pol12,kru15,mar17}.
Here we closely follow a similar analysis applied recently to $^{48}$Ca \cite{bir17}.
As in Ref.~\cite{bir17} we neglect M1 contributions since the M1 strength in $^{40}$Ca is concentrated in a single transition at 10.318 MeV \cite{gro79}.
Besides Coulomb excitation of the GDR and isoscalar E0, E2, and E3 collective excitations, we allow for a nuclear background (dominated by quasifree scattering), whose angular distribution is assumed to be constant \cite{hau91}.  

%
The resulting decomposition is presented in Fig.~\ref{fig:mda} for the $0^\circ$ spectrum covering the full angular acceptance.
Contributions from E0, E2 and E3 modes are small consistent with findings in $^{48}$Ca \cite{bir17} and in heavier nuclei \cite{tam11,pol12,kru15,mar17}.
The MDA confirms the excitation of the GDR peak by relativistic Coulomb excitation.
The magnitude of the nuclear background of about 2 mb/(sr MeV) is again consistent with the MDA results in heavier nuclei and with other measurements at similar incident proton energies \cite{hau91,bak97}. 

\section{RPA calculations} 

For comparison with the experimental measurements, theoretical $B(E1)$ strength distributions were calculated in the random phase approximation (RPA) starting from axially deformed Hartree-Fock (HF) ground states and using explicit angular-momentum projection techniques. 
Both the HF and the RPA calculations use the same realistic nucleon-nucleon interaction derived from the Argonne V18 potential by a unitary transformation in the framework of the Unitary Correlation Operator Method (UCOM) \cite{fel98,rot10} and are supplemented by a phenomenological three-nucleon contact interaction. 
This Hamiltonian was introduced and tested in Ref.~\cite{gun10}  for ground-state observables and applied for RPA calculations in closed-shell nuclei in Ref.~\cite{gun13} (we use the version labeled `S-UCOM(SRG)'). 
All calculations were performed in a harmonic-oscillator single-particle basis covering 15 oscillator shells. 
Further details on the deformed RPA approach employed in this work can be found in Ref.~\cite{erl14}.
For $^{24}$Mg and $^{28}$Si, theoretical results are also available  from a selfconsistent axially-symmetric deformed
Hartree-Fock-Bogolyubov (HFB) plus quasiparticle RPA (QRPA) calculation with the D1S Gogny force \cite{per08}.

The resulting theoretical strengths are shown in the middle and bottom rows of Fig.~\ref{fig:data}, respectively.
The location of the peaks in the calculated spectra of $^{28}$Si, $^{32}$S, and $^{40}$Ca (middle row) is roughly consistent with what is seen experimentally (top row), i.e.\ the predicted spreading of the strength in qualitative agreement with the data.
This is somewhat surprising since e.g.\ the fragmentation of the GQR -- albeit in heavier nuclei \cite{she09} -- needs a description in terms of a second-RPA approach.
For $^{24}$Mg, a discrepancy is observed at higher excitation energies, where a cumulation of strength is predicted  around 22 MeV without an experimental counterpart.
Both, cluster models \cite{kim12} and density functional approaches \cite{yao10} indicate the presence of triaxial deformations in $^{24}$Mg, which are not accounted for by our calculation and could explain the larger deviation we observe for this nucleus. 
The calculations of Ref.~\cite{per08} for $^{24}$Mg and $^{28}$Si (bottom row) roughly reproduce the centroids but show too little spreading of the strength. 

The model calculations do not include the continuum and so the strength distributions consist of discrete transitions.
For the subsequent wavelet analysis the theoretical distributions hve been folded with a Gaussian of a width corresponding to the experimental resolution so that the low-scale cutoff in the wavelet power spectra (see below) matched the experimental data.
These (Q)RPA strength distributions and the experimental ones are now analyzed with the wavelet method.

\section{Wavelet Analysis}

\begin{figure}[t]
\includegraphics[width=8.5cm]{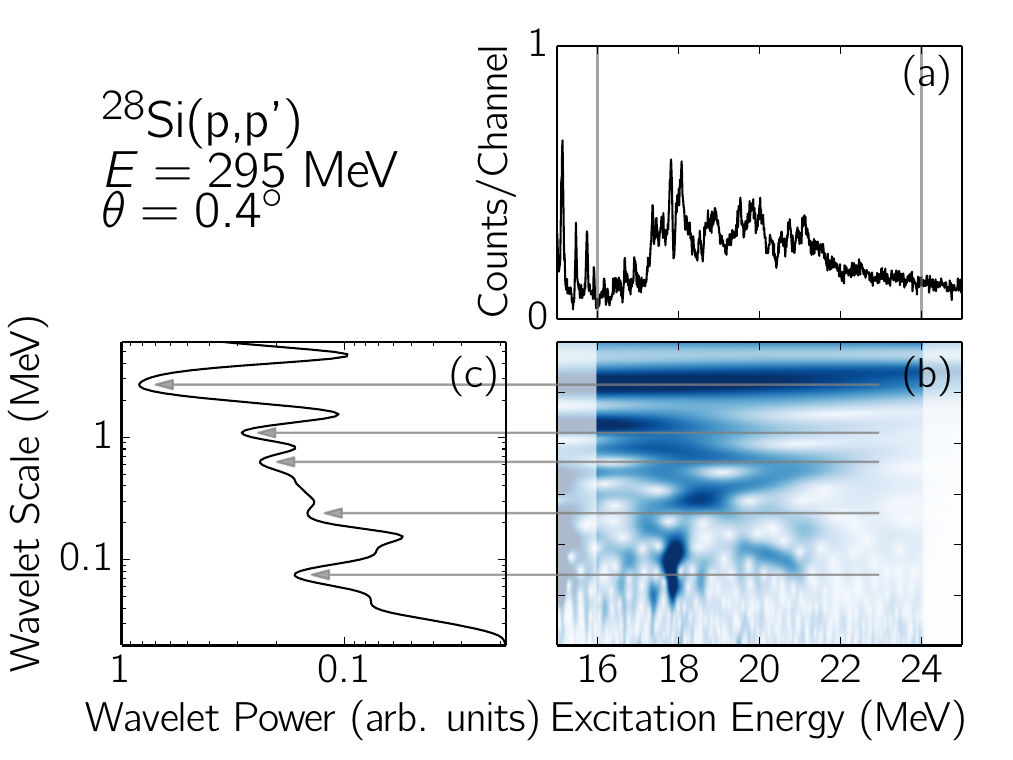}
\caption{Example of the wavelet analysis. 
(a) Experimental spectrum of the GDR in $^{28}$Si from the $(p,p^\prime)$ reaction. 
(b) Square of the wavelet coefficient $C$ [Eq.~(\ref{eq:wc})] as a function of $E_x$ and wavelet scale. 
The highlighted area ($16-24$ MeV) is selected for projection onto the scale axis. 
(c) Wavelet power spectrum. 
The peaks quantiatively characterize locations and widths of the fine structure.}
\label{fig:cwt}
\end{figure}

The wavelet analysis of the measured spectra is illustrated by the example of the $^{28}$Si($p,p^\prime$) data [Fig.~\ref{fig:cwt}(a)]. 
It proceeds via the calculation of a wavelet coefficient $C$ from the measured cross sections $\sigma(E)$ (expressed in Counts/channel)
\begin{equation}
C_i(\delta E)\equiv C(\delta E,E_i)=\frac{1}{\sqrt{\delta E}} \int \sigma(E) \Psi^*\left(\frac{E_i-E}{\delta E}\right) dE,
\label{eq:wc}
\end{equation}
where $E_i$ is the excitation energy of channel $i$, $\delta E$ the wavelet scale, and $\Psi$ the wavelet function. 
Here, the complex Morlet wavelet 
\begin{equation}
\Psi(x)=\pi^{-1/4}\,e^{ik_0x}\,e^{-x^2/2},
\label{eq:wf}
\end{equation}
with $k_0=5$ is employed, which provides optimum balance between resolution of excitation energy and energy scale for the present application (see, e.g., also Ref.~\cite{usm16}).
The wavelet decomposition is performed over the whole spectrum with reflective boundary conditions and a region of interest corresponding to the bulk of the GDR strength.  
The squares of the wavelet coefficients, representing a measure of the strength of structures resolved by the wavelets, are displayed in Fig.~\ref{fig:cwt}(b).

Because of possible contributions to the spectra from spin-$M1$ excitations at lower excitation energies \cite{mat15,hey10}, further analysis is restricted to the highlighted area ($16-24$ MeV).
The projected power spectrum
\begin{equation}
P_w(\delta E)=\frac1N \sum_{i=i_1}^{i_2}|C_i(\delta E)C^*_i(\delta E)|,
\label{eq:power}
\end{equation}
where $i_1$ and $i_2$ indicate the boundaries of the region of interest, is shown in the lower left hand panel (c).
Peaks of strength in this power spectrum are associated with characteristic scales of the structures in the region of the GDR. 
The power spectrum is normalized to the spectral variance in order to facilitate comparison between different nuclei and with theoretical results. 
The analysis of the fluctuations, if represented as a power, characterizes the variance of the series under consideration. 
The Fourier transform preserves the variance of the signal and the CWT does as well (at least approximately) since it is a convolution. 
Thus, a normalization to the variance facilitates a comparison of powers deduced from the various spectra despite the absence of an absolute scale.      

\begin{figure}[t]
\includegraphics[width=8.5cm]{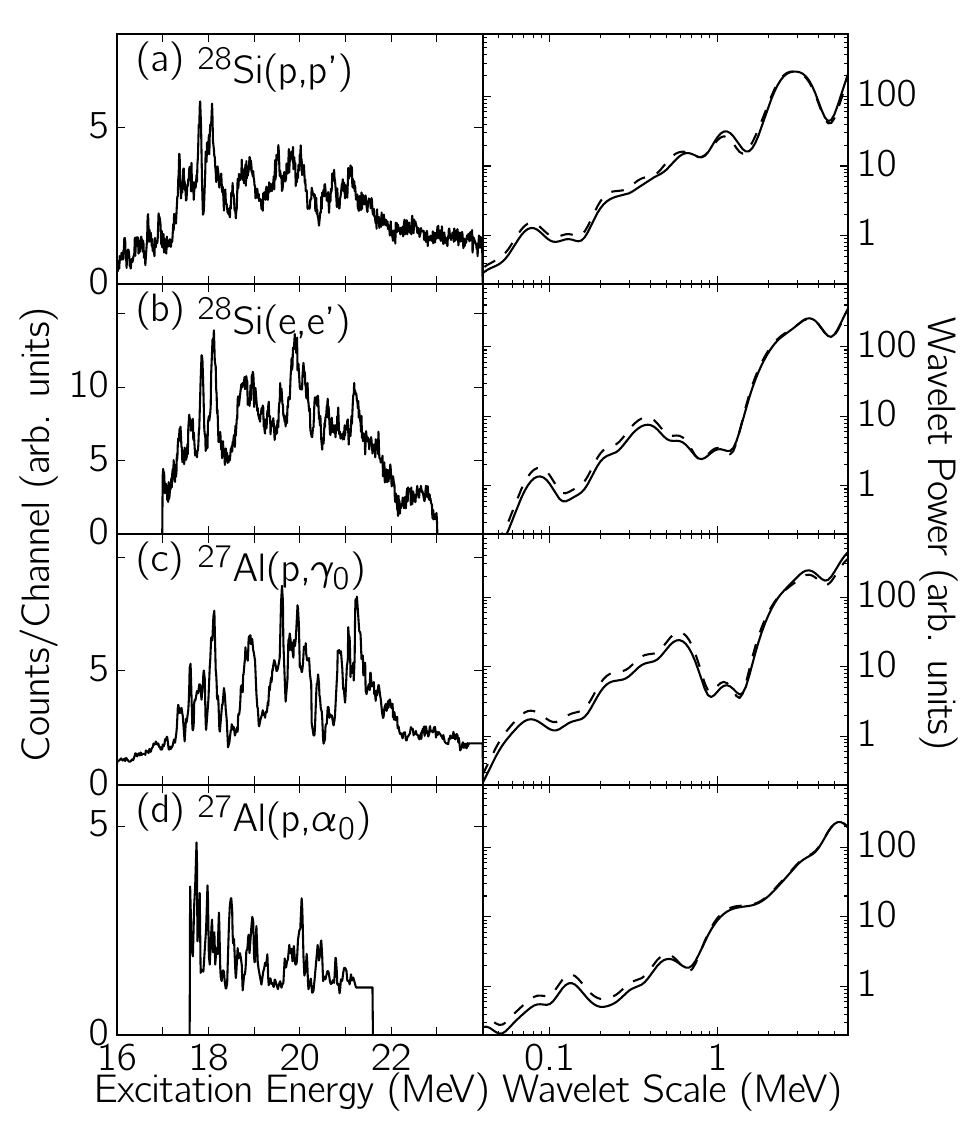}
\caption{Left: Experimental data from different high-resolution experiments populating the GDR in $^{28}$Si.
Right: Power spectra from the wavelet analysis summed over excitation energy regions $16-24$ MeV (solid line) and $17 -23$ MeV (dashed line).}
\label{fig:all28}
\end{figure}

For the case of $^{28}$Si there are several sets of high resolution data available in the literature shown in the l.h.s.\ of Fig.~\ref{fig:all28}: (a) present work, (b) the $^{28}$Si$(e,e')$ reaction \cite{fri81,ric85}, (c) the  $^{27}$Al$(p,\gamma)$ reaction \cite{sin65}, and (d) the $^{27}$Al$(p,\alpha_0)$ reaction \cite{law65,put68}.
It is expected that reactions (a)-(c) predominantly excite the GDR. 
Reaction (d) favors isospin $T = 0$ states in $^{28}$Si and is therefore not selective towards $1^-$ levels but may provide a possible window into more general origins of fine structure. 
These data were analyzed in the same way and the resulting wavelet power plots are displayed on the r.h.s.\ of Fig.~\ref{fig:all28}. 
Not only can corresponding structures be located in the experimental spectra, but also in the wavelet power plots there is a good agreement demonstrating the utility and reliability of the wavelet method.
This is particularly evident comparing the $(p,p^\prime)$, $(p,\gamma)$, and $(e,e^\prime)$ reactions.
The similarities of the results underline that the structures extracted with the wavelet analysis are indeed intrinsic features of the nuclei involved.
\begin{figure*}
\includegraphics[width=16cm]{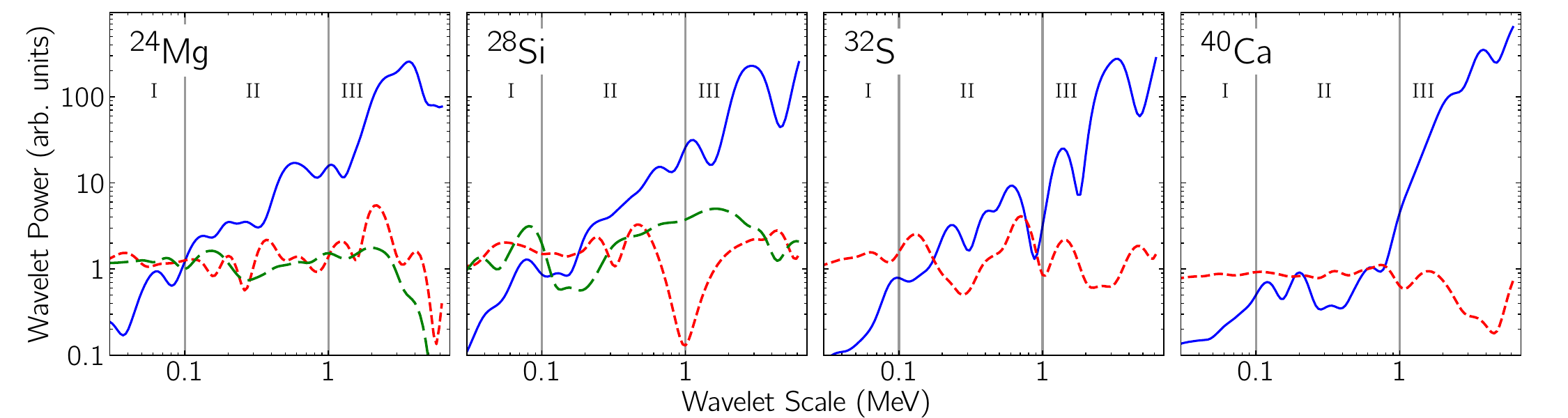}
\caption{Wavelet power spectra of the GDR in  $^{24}$Mg,  $^{28}$Si,  $^{32}$S, and  $^{40}$Ca from the experimental and theoretical data of Fig.~\ref{fig:data}.
Blue solid lines: experiment. Short-dashed red lines: HF+RPA calculatons with S-UCOM(SRG) interaction \cite{erl14}. Long-dashed green lines: HFB+QRPA calculations with D1S Gogny interaction \cite{per08}.}
\label{fig:power}
\end{figure*}

We have also investigated the sensitivity to the energy window chosen for the wavelet analysis.
The green dashed lines in Fig.~\ref{fig:all28} show the power spectra resulting from a summation of the wavelet coefficients over the energy region $17-23$ MeV rather than $16-24$ MeV.
As one can see, the differences are very small.

\section{Discussion}

Table \ref{tab:scales} summarizes the significant scales in $^{28}$Si in the wavelet power spectra for all experimental data and theoretical calculations. 
These can be grouped in three classes: In all the experimental spectra there is a scale (Class I) at approximately 80 keV.
This is similar to the average level width due to Ericson fluctuations in $^{28}$Si \cite{eri66,ebe72} and we tentatively follow this identification.
It conforms with the absence of a corresponding scale in the theoretical results.
We note that the wavelet analysis of the strength distribution from Ref.~\cite{per08} also shows a scale at about 80 keV.
However, the two-dimensional correlation analog to Fig.~\ref{fig:cwt}(b) traces its origin back to a series of transitions localized in a small energy interval of $E_{\rm x} \simeq 20 -21$ MeV.
Thus, the scale does not represent a genuine feature of the GDR.
\begin{table}[b]
 \caption{Characteristic scales of the fine structure of the GDR in  $^{28}$Si from different experiments and the theeoretical results of Refs.~\cite{erl14} and \cite{per08}.
Scales are divided into three classes (see text).}
\label{tab:scales}
 \begin{ruledtabular}
  \begin{tabular}{lcccc}
   Spectrum & Ref. &  \multicolumn{3}{c}{Scales (MeV)} \\
&  & \multicolumn{1}{c}{Class I}&\multicolumn{1}{c}{Class II}&\multicolumn{1}{c}{Class III}\\
\hline
$^{28}$Si$(p,p^\prime)$ & present & 0.08 &\phantom{0.00} 0.23 0.36 0.59 & 1.0 \phantom{0}2.9 \\
$^{28}$Si$(e,e')$ & \cite{fri81,ric85} & 0.08 &  0.23 0.36 & 1.0 \phantom{0}3.3 \\
$^{27}$Al$(p,\gamma)$ & \cite{sin65} &  0.08 & 0.14 0.24 0.38 0.59 & 1.1 \phantom{0}3.2 \\
$^{27}$Al$(p,\alpha)$ & \cite{law65,put68} & 0.09 & 0.12\phantom{0.00 } 0.40\phantom{0.00,} & 1.0 \phantom{0}2.6 \\
\hline
RPA & \cite{erl14} & &  0.23 0.44 &\phantom{0.00, }2.1 \\  
QRPA   & \cite{per08} & \phantom{0}(0.08)\footnote{Not a genuine scale of the GDR. see text.} & \phantom{0.00} \phantom{0.00} 0.42 0.77 &  \phantom{0.00} 1.6 \\
\hline
\end{tabular}
\end{ruledtabular}
\end{table}

At  larger energies scales similar in energy to those in $^{28}$Si$(p,p^\prime)$ are seen in the $^{28}$Si$(e,e^\prime)$ data and in the $^{27}$Al$(p,\gamma)$ data. 
The $^{27}$Al$(p,\alpha)$ data shows similar numbers of scales but with slightly shifted energy. 
These are denoted Class II and Class III, where Class III scales are large scales associated with the spread of the distribution of strength while Class II are intermediate scales in the region 100 keV to 1 MeV\footnote{We note that the choice 100 keV and 1 MeV as borders to distinguish between the different classes is somewhat arbitrary.
The values were chosen to facilitate easy comparison to previous studies of the GDR in $^{208}$Pb \cite{pol14} and of the GQR in many nuclei \cite{she04,she09}.}. 
Similar results are obtained for the other nuclei chosen for study, $^{24}$Mg, $^{32}$S and $^{40}$Ca as illustrated in Fig.~\ref{fig:power} although in the case of $^{40}$Ca the Class II scales are rather weak and the power spectrum is dominated by Class III.

The power spectra extracted from the data and from the theoretical methods are compared in Fig.~\ref{fig:power},
The power spectra from theory reproduce features of the experimental data like the common observation of a scale around 100 keV and a typical number of $3-4$ scales between 100 keV and 1 MeV. 
However, the overall rise of power at larger scales in the data is not observed. 
To some extent this is due to the ability of the wavelet analysis to resolve features both in spacing and in shape. 
The Morlet wavelet yields local wavelength information because of its oscillatory shape but it also functions as a generalized second derivative operator giving an enhanced signal for the broad bell-shaped distribution of GDR strength in the experimental spectra.
Thus, large scales with high strength observed in the experimental data are not replicated in the theoretical analysis because neither the continuum nor the damping due to coupling to many-particle many-hole configurations is included in the calculations.
Convolution of the theoretical strengths with a Lorentzian curve of suitable width broadens the lines and indeed enhances the large-scale wavelet power but is artificial in the absence of knowledge of the true widths.

The observation of class II scales comparable to those seen in the analysis of the experimental  data demonstrate that already at the RPA level in our realistic calculations there is considerable fragmentation of the strength.
The origin of this fragmentation in the theoretical spectrum is the deformation of the nucleus most likely driven by the strong $\alpha$ clustering (for experimental evidence see e.g.\ \cite{fre17}) in these nuclei \cite{erl14}.
This suggests that a prime source of the fine structure in the GDR in light nuclei is deformation rather than the
coupling to surface vibrations invoked for the GQR in heavier nuclei \cite{she04,she09}. 
This observation is supported by the case of the closed shell nucleus $^{40}$Ca where both experiment  and theory exhibit only weak fine structures and correspondingly little wavelet power of Class I and II.

\section{Conclusions}
We have available, for the first time, a set of high resolution data for the GDR region of $N=Z$ nuclei in the $sd$ shell, together with RPA calculations performed on top of a deformed ground state with a realistic nucleon-nucleon interaction. 
This enables us to investigate the long-standing question of the origin of fine structure of the GDR in nuclei in this mass region.
A wavelet analysis permits to extract scales characterizing this fine structure and the results for different reactions exciting the GDR in $^{28}$Si shows that good consistency is achieved. 
Comparisons  between experimental data and the RPA calculations suggest that fine structure at the level of a few hundred keV (class II scales) results mainly from the deformation of the nuclei driven by $\alpha$ clustering. 
This is in sharp contrast to the case of the GQR where coupling to $2p2h$ states is the main source of characteristic scales in the region 100 keV - 1 MeV \cite{she04,she09}.   

\begin{acknowledgments}
We are indebted to S.~P\'{e}ru for providing us with the numerical results of Ref.~\cite{per08}.
This work was supported the DFG under contract SFB 1245, by the South African NRF, and by JSPS KAKENHI grant number JP25105509.

\end{acknowledgments}

\end{document}